\documentstyle[12pt]{article}

\def\beqra{\begin{eqnarray}}
\def\eeqra{\end{eqnarray}}
\def\beqast{\begin{eqnarray*}}
\def\eeqast{\end{eqnarray*}}
\def\be{\begin{enumerate}}
\def\ee{\end{enumerate}}

\def\beq{\begin{equation}}
\def\eeq{\end{equation}}

\begin{document}

%\today

%\hfill{LBL 33016}

\vspace{24pt}

\begin{center}

{\large{\bf Generation Structure and Proton Decay Problem                                  }}

\vspace{36pt}
K. Koike

\vspace{4pt}
Department of Natural Science \\ Kagawa University
\\ Takamatsu, 760-8522 Japan

\vspace{30pt}
{\bf Abstract}

\end{center}

\begin{minipage}{4.75in}

 Possible generation structure is investigated. We considere alternative
possibility of combination of leptons and quarks to compose each 
generation, where the combination of $(\nu_{\tau},{\tau})$ and $(u,d)$
compose a generation. Our model exhibits the hierarcy structure
after sea-saw mechanism provived that possible existence of realistic 
right-handed neutrinos are taken into account. Then, in the GUTs model 
based on our scheme, possible inter-generation properties are examined 
together with the classification of gauge bosons. 
It is shown that if our scheme is realized, the proton decay exhibits 
different mode from ordinary one.
\end{minipage}

\section{Introduction}

 The generation structure observed in low energy region seems to 
exhibit an important suggestion to disclose fundamental existence-form of 
matter.\cite{rf:Maki,rf:KK}. It will refrect probably the structure 
of nature in deeper level. In the standard comprehension, the generation 
is simply arranged from light to large mass acording to the historical discovery 
order which is due to the energy frontier. According to this custom, the 
combination of $(\nu_{e},{e})$ and $(u,d)$ compose the 
first generation. It should be noted that, however, the essential reason
to relate leptons and quarks is to satisfy the anomaly free 
condition, and no restriction is yet known to what lepton must compose a 
generation together with a certain specific quark. Then, it is worthwhile to 
investigate another possibe generation model as a step to disclose the origin 
of generation structure. This paper is concerning to this problem.

 In this paper, we will propose an alternative model of generation combination of
leptons and quarks. In our model, the combination of $(\nu_{\tau},{\tau})$ 
and $(u,d)$ compose a generation. It is shown that our model exhibits the new
hierarcy structure after sea-saw mechanism provived that the right-handed 
neutrinos with appropriate mass are taken into account. 
Then, the generation structure is investigated in the GUTs scheme. In what stage 
of symmetry breaking of GUTs, is the generation structure appear? We will 
suppose that a partial GUT structure\cite{rf:pGUT} in each generation in the 
series of breaking of GUTs symmetry.
This picture leads to possible classification of the gauge bosons appearing
in GUTs scheme. 
We will examine possible properties of gauge bosons both within one generation 
and among inter-generation.
Especially, our scheme exhibits the different mode of proton decay in the
framework of ordinary GUTs scheme without super symmetry.

% In investigating to that 
%structure, neutrino problem may be situated at the capstone. It seems 
%that the see-saw mechanism is the most prominent one to explain the 
%smallness of neutrino mass, where its smallness is reduced to 
%largeness of the right-handed Mayorana neutrino mass which is often 
%assumed to be the order of GUTs mass. In the GUTs scale, a new 
%physics beyond to  our present prediction may appear,
%and it is not clear whether the right-handed Mayorana neutrino with
%GUTs mass can be treated as ordinary particle.
% On the other hand, if we assume that the right-handed Mayorana 
%neutrino possesses mass 
%below GUTs scale, and can be completely treated as ordinary particles
%described by ordinary gauge field theory,
%possible alternative generation structure will be implied.
 
%In this paper, we will investigate possible new generation structure
%on the basis of viewpoint suggested by the see-saw mechanism, and 
%discuss on a few typical new features due to this scheme.
% This paper is concerning to this problem.

%It is often supposed, behind of the structure,
%the existence of GUTs scheme

% The GUTs mass order may imply, beyond to its original meaning in the
%gauge field theory, the appearance of a new mechanism of physics. 
%If this is true, we may not treat the right-handed Mayorana 
%neutrino as usual particles. 

\section{Possible scheme of generation Structure}

 Our model of the new generation combination is represented as,

\beq
\begin{array}{lll}
\\
I&~~~~~{{\nu_{{\tau}L}} \choose {\tau}_L}~~& {{u_L} \choose {d_L}}\\
\\
II&~~~~~{{\nu_{{\mu}L}} \choose {\mu}_L}~~ &{{c_L} \choose {s_L}}\\
\\
III&~~~~~{{\nu_{{e}L}} \choose {e}_L}~~ &{{t_L} \choose {b_L}}\\\
\end{array}
\label{eq:newgene0}
\eeq
where in Eq.(\ref{eq:newgene0}), we have written only the left-hand commponents.
We will suppose possible existence of GUTs structure behind of the generation 
structure,
that is, as a result of spontaneous symmmetry breaking, our generation 
structure is realized. The mass spectrum of leptons and quarks will be 
realized as a result of symmetry breaking with Higgs mechanism.

\section{See-saw mechanism and possible new hierarchy structure}

In our model, so far as we are restricted in well known particles,
a hierarchy structure  with simple form seems to not appear.
It should be noted that, however, possible new hierarcy structure will
appear provided that the existence of right-handed Majorana neutrinos 
are taken into account together with the sea-saw mechanism. Then, we 
will scketch this problem.

\subsection{See-saw mechanism of neutrino mass splitting}

\indent
 For definiteness, we will make a quick review of see-saw mechanism of
neutrino mass generation.\cite{rf:GRSY}
As a basis of construction of our scheme, let
us consider the D-M (Dirac-Mayorana) mass term 
\cite{rf:Pon},\cite{rf:Bilen} in the simplest case of
one generation labeled by the generation suffix i.
We have

\begin{eqnarray}
{\cal{L}}^{D-M}& = &
-\frac{1}{2} m_{iL} \overline{(\nu_{iL})^c} \nu_{iL}
-m_{iD}\bar{\nu}_{iR}\nu_{iL}
-\frac{1}{2} m_{iR} \bar{\nu}_{iR} (\nu_{iR})^c~~+~~h.c.
                  \nonumber\\
               & = &-\frac{1}{2}
\overline{{(\nu_{iL})^c} \choose {\nu_{iR}}}
{}~M~{{\nu_{iL}} \choose {\nu_{iR}^c}}
{}~~+~~h.c.
\label{eq:DM}
\end{eqnarray}

\noindent Here

\beq
M = \left( \begin{array}{cc}
m_{iL} & m_{iD} \\
m_{iD} & m_{iR}
\end{array} \right),
\label{eq:MATm}
\eeq

\noindent $m_{iL}, m_{iD}, m_{iR} $ are parameters. For a symmetrical 
matrix M we have

\beq
M=U~ m~ U^{\dag},
\label{eq:DIA}
\eeq

\noindent where $U^{\dag}U=1 $, $m_{jk}=m_j \delta_{jk} $. From 
Eq.(\ref{eq:DM}) and
Eq.(\ref{eq:DIA}) we have

\beq
{\cal{L}}^{D-M}=-\frac{1}{2} \sum_{\alpha=1}^{2}
 m_{i{\alpha}}{\bar{\chi}}_{{i\alpha}} \chi_{i{\alpha}} ,
\label{eq:dia2}
\eeq

\noindent where

\begin{eqnarray}
\nu_{iL}~~  =~~~{\cos{\theta_i}}\chi_{i1L} & + & {\sin{\theta_i}}\chi_{i2L}
\nonumber\\
(\nu_{iR})^c  =  {-\sin{\theta_i}}\chi_{i1L} & + & {\cos{\theta_i}}\chi_{i2L}.
\label{eq:MIX}
\end{eqnarray}

\noindent Here $\chi_{i1} $ and $\chi_{i2} $ are fields of Majorana
 neutrinos with masses $ m_{is}~(small), m_{iB}~(Big) $, respectively.
  The masses $m_{is}, m_{iB} $ and the mixing angle
$\theta_i $ are connected to the parameters $m_{iL}, m_{iD} $ 
and $m_{iR} $ by the relations

\begin{eqnarray}
m_{is}~ & = & \frac{1}{2}~ {\left|{m_{iR}~ +~m_{iL}~-~a_i} \right|}
\nonumber\\
m_{iB} & = & \frac{1}{2}~ {\left|{m_{iR}~ +~m_{iL}~+~a_i} \right|}
\nonumber\\
\sin{2\theta_i} & = & \frac{2m_{iD}}{a_i},~~~~\cos{2\theta_i}
=\frac{m_{iR}-m_{iL}}{a_i}
\label{eq:MIX2}
\end{eqnarray}

\noindent where

\beq
a_i=\sqrt{(m_{iR}-m_{iL})^2~+~4{m_{iD}^2}}
\label{eq:a}
\eeq

It should be noted that the relations Eq.(\ref{eq:MIX2}) are exact ones. 
Let us assume now that

\beq
m_{iL}=0,~ m_{iD}\simeq{ m_{iF}},~m_{iR}\gg{ m_{iF}} ,
\label{eq:mass-sb1}
\eeq

\noindent where $m_{iF} $ is the typical mass of the leptons and quarks of
the generation labeled by suffix i. From Eq.(\ref{eq:MIX2})  
we have

\beq
m_{is}\simeq\frac{m_{iF}^2}{m_{iR}},~~m_{iB}\simeq{m_{iR}} ,
{}~~\theta_i\simeq{\frac{m_{iD}}{m_{iR}}}
\label{eq:mass-sb2}
\eeq

\noindent
 Thus, if the conditions Eq.(\ref{eq:mass-sb1})  are satisfied,
the particles with definite masses are split to a very light Majorana
neutrino with mass
$m_{is} \ll m_{iF} $ and a very  heavy Majorana particle with mass
$m_{iB}\simeq m_{iR} $. The current neutrino field $\nu_{iL} $ practically
coincides with
 $\chi_{i1L} $ and $ \chi_{i2}\simeq {\nu_{iR}~+~(\nu_{iR})^c} $
, because $\theta_i$ is extremely small.

That is , we have assumed such scheme that in
D-M mass term Dirac masses  are of order of usual fermion masses, 
the right-handed Majorana masses, responsible for lepton numbers
violation, are  extremely large  and the left-handed 
Majorana masses are equal zero. In such a scheme  neutrinos are Majorana
particles with masses much smaller than masses of the other fermions.
The predictions of neutrino masses depend on the value of the
$m_{iR}$ mass.
The value of $m_{iR}$ is often assumed that 
$m_{iR}=M_{GUT}$, where $M_{GUT} $ is grand unification scale.
Though this value  depends on the model, a typical one is
$m_{iR}\simeq {10^{19}} $ GeV (Planck mass). In the $M_{GUT}$ region, 
the ordinary particle picture may be drastically changed.
\par
It should be emphasized that, however, there is no definite reason 
why the mass of  $m_{iR}$ should be $M_{GUT}$.  It is also possible 
that though the mass of  $m_{iR}$
is very huge it is  below the Planch mass and  possessing the picture of
ordinary particle. In such case,  possibility of realization of new 
generation hierarchy will be implied, which is discussed in the next 
section.

\subsection{Possible hierarchy structure based on the right-handed Majorana neutrino  }

  The "standard" structure of generation is composed of each leptons 
and quarks, which generation number is labeled according to the sequence 
of their historical discovery, depending on the energy frontier. 
It should be emphasized that, in the present stage, the only compelling 
reason to relate leptons and quarks is the anomaly free condition, and no 
essential principle to compose the generation is ever unknown.\cite{rf:GUTs} 
In our generation scheme given in Eq.(\ref{eq:newgene0}), however, a simple
hierarchy structure will appear provided that the right-handed Mayorana neutrinos
are taken into account.

%Then, regarding the conventional rule to classify the generation according to 
%the sequence of magnitude of the total masses belonging to the same suffix as 
%the fundamental rule which should be founded on the deeper level, we will 
%investigate alternative possible generation structure.

%\par
If the right-handed Majorana neutrinos are realistic particles below the 
GUT mass and responsible to the see-saw mechanism, its existence will 
affect to the above mentioned classification procedure of the generations.
%\par
From Eq.(\ref{eq:mass-sb2}), $m_{iR}$ is expressed as,
\beq
m_{iR}\simeq\frac{m_{iF}^2}{m_{is}},
\label{eq:mass-sb3}
\eeq
where we should remember to the fact that the current neutrino field $\nu_{iL} $ 
practically coincides with  $\chi_{i1L} $ with the mass $m_{is}$.
Though the values of neutrino mass has not yet been established,
\cite{rf:nu-mix} a prominent
possibility is that the mass of neutrinos exhibits an extreme hierarical
structure%\cite{rf:Tanimo},
\beq
m_{\nu_e}\ll{m_{\nu_\mu}}\ll{m_{\nu_\tau}}.
\label{eq:mass-inequal}
\eeq
If the degree of the mass difference in Eq.(\ref{eq:mass-inequal}) is 
appropriate magnitude, the new hierarical structure will be realize.
That is, if the degree of mass difference of neutrinos is larger than that of
the representative mass of each generation $m_{iF}$, the magnitude of total 
mass of i-th neutrinos,
i.e. $m_{\nu_i} + m_{iR}$, will become in  reverse order to label number i,
and this magnitude will dominate the total lepton masses belonging to i-th 
label.

Then, we will find the hierarical structure of "from light to heavy" particles
in the following generation cllasification.
%Then, provided that we adopt the custom to nominate the generation number 
%"from light to heavy" particles, we will conclude  the following generation 
%classification, 

%\vfill\eject

\beq
\begin{array}{lll}
\\
I&~~(\nu_{1R})^c~~~{{\nu_{{\tau}L}} \choose {\tau}_L}~~& {{u_L} \choose {d_L}}\\
\\
II&~~(\nu_{2R})^c~~~{{\nu_{{\mu}L}} \choose {\mu}_L}~~ &{{c_L} \choose {s_L}}\\
\\
III&~~(\nu_{3R})^c~~~{{\nu_{{e}L}} \choose {e}_L}~~ &{{t_L} \choose {b_L}}\\\
\end{array}
\label{eq:newgene}
\eeq
where we have written only the left-handed components.
That is, taking into account of the existence of realistic particles $\nu_{iR}$,
our proposal for new generation structure is different from the ordinary
one. The characteristic feature is that the leptons belonging to the first and 
the third generations are exchanged. It should be noted that this result is 
caused by the existence of realistic $\nu_{iR}$ and the see-saw mechanism.

\section{"Generational" and "Inter-generational" gauge bosons in GUTs
and proton decay problem}

The generation structure is the important feature observed in low energy 
region. This structure will still possess any meaning in high-energy region 
below GUTs scale. We will suppose that in the symmetry breaking chain of GUT,
the single-generation unification structure appear prior to the apperence
of each generation structure.\cite{rf:pGUT}
This situation leads to the viewpoint to distinguish the gauge bosons appearing
in GUTs containing all generations. That is, the gauge bosons common to all 
generations, and ones which connect particles belonging to different generations.
In SU(5) GUT, only the former type of gauge bosons, $W, Z, A$ and $X, Y$ , appear. 
It should be emphasized that though possible gauge group is not restricted to 
SU(5), we can see the characteristic feature of GUT in this model.  We see that 
the type of gauge bosons X and Y appearing in SU(5) model will appear in any 
single-generation unification structure. We will call this type of gauge bosons 
as "generational gauge bosons"\cite{rf:Abe-lq2}. In the GUTs structure 
containing all generations, new gauge bosons connecting particles
belonging to the different generations appear generally, and we will call
them as "inter-generational gauge bosons".\cite{rf:flavor-c} 
Our viewpoint means that the 
contribution of the "inter-generational gauge boson" should be more 
suppressed than that of the "generational gauge bosons" in low and 
intermediate energy region below the GUTs scale.

If we take this viewpoint, our model of new generation structure 
lead to different feature of proton decay mode in GUTs.
Our scheme predict the proton decay mode due to the generational gauge 
bosons $X$ and $Y$

\beq
P~\rightarrow~\tau^+~M_0
\label{eq:decay-mode}
\eeq

\noindent
instead of the well known mode
\beq
P~\rightarrow~e^+~M_0
\label{eq:decay-mode-e}
\eeq

\noindent
where $M_0$ represents $\pi^0,\rho_0,\omega,\eta,\pi^+\pi^-\cdots$.
It should be noted that the process in Eq.(\ref{eq:decay-mode}) is
forbidden by Q value.
The inter-generational gauge bosons may cause the process in 
Eq.(\ref{eq:decay-mode-e}), however, this process will be extremely
suppressed in low and intermediate energy region below GUTs scale.
 The other mode is

\beq
P~\rightarrow~{\overline{\nu}}_\tau~M^+
\label{eq:decay-mode-nu}
\eeq

\noindent
with $\pi^+,\rho^+,\pi^+\pi^0\cdots$, and this process is allowed
in our scheme.

Thus, the proton decay in our model is different from the "standard"
model of generation structure. That is, the well-known difficulty 
so-called "proton decay problem" is consistent to our scheme. It 
should be emphasized that when the proton decay is actually observed 
it will be made clear whether our scheme is realized or not.

\section{Discussion}

  In this paper, we have proposed a possible new structure of generation 
combination of leptons and quarks. It is discussed that our scheme exhibits
a hierarchial structure through the see-saw mechanism provided that the 
right-handed Mayorana neutrinos are  actual particles with approapriate mass.
It should be noted that this structure appears as a result of the breakdown of 
a certain GUT structure and successive see-saw mechanism. The profound meaning
of the appearence of new hierarchial structure via above two step process 
may be clarified in possible dynamics of sub-structure or in the process of
disclosing futher fundamental existence form of matter.\cite{rf:Maki,rf:KK}

We have classified the gauge bosons appearing in GUTs, 
that is the "generational gauge bosons" and the "inter-generational gauge 
bosons". Our model gives some important predictions to so-called "proton 
decay problem". When the proton decay is actually observed and the neutrino
masses are established, it will be made clear whether our scheme is realized 
or not.  

We have taken in this 
paper so-called constructivity-like approach. That is, supposing the 
existence of GUTs or partial GUT-like structure behind our discussion,
the final form of it doesn't appear explicitly.
In order to give further predictions, some specific assumptions 
and introduction to many parameters is inevitable in the present stage, 
then we have restricted ourselves to proposal of the framework of model. 
In this paper, we have  investigated in the framework
of ordinary GUT scheme without supersymmetry because the generation 
structure will be prior to a possible higher symmetry. The 
investigation in the framework of SUSY GUT is further problem.

\vfill\eject

%\section{Remarks}

%notation

%\parskip=5mm %$m_i^{(s)}$ %\par %$m_i^{(b)}$

%\subsection{Fractions}

%\section*{References}

%\section*{Acknowledgements}
%We would like to thank ...........

%\appendix

%\section{Second Appendix}


\begin{thebibliography}{99}
\bibitem{rf:pGUT}
That is, in a series of breaking of GUTs symmetry, it appears a partial GUT 
structure such as the SU(5) GUT in each generation.
\bibitem{rf:GUTs}
The mass relation in GUTs means 
merely indirect condition. If new structure of generation is established,
the modified GUTs relation will be constructed on the basis of it.
\bibitem{rf:nu-mix}
  In order to determine the neutrino masses, the mixing structure should
be completely clarified. In the present stage, it will be premature to find
the definite masses of neutrinos.
\bibitem{rf:SU(5)type}
Though possible 
group is not restricted to SU(5), it should be emphasized that we can see the 
characteristic feature of GUT in this model.  We see that the type of gauge bosons 
X and Y appearing in SU(5) model will appear in any single-generation unification
structure.
\bibitem{rf:flavor-c}
 "Inter-generational gauge bosons" will generally cause 
 flavor-changinng effects.
\end{thebibliography}
\end{document}